\documentclass[conference]{IEEEtran}
\IEEEoverridecommandlockouts
\usepackage{cite}
\usepackage{amsmath,amssymb,amsfonts}
\usepackage{algorithmic}
\usepackage{graphicx}
\usepackage{textcomp}
\usepackage{xcolor}
\def\BibTeX{{\rm B\kern-.05em{\sc i\kern-.025em b}\kern-.08em
    T\kern-.1667em\lower.7ex\hbox{E}\kern-.125emX}}
\usepackage{graphicx}
\usepackage{subcaption}
\usepackage{booktabs}
\usepackage{tikz}
\usetikzlibrary{positioning, arrows.meta}
\usepackage{balance}

\begin{document}

\title{Agentic AI for Improving Precision in Identifying Contributions to Sustainable Development Goals\\
\thanks{This project was made possible in part by the Institute of Museum and Library Services (LG-256638-OLS-24).}
}

\author{
    \IEEEauthorblockN{1\textsuperscript{st} William A. Ingram}
    \IEEEauthorblockA{
        \textit{University Libraries} \\
        \textit{Virginia Tech}\\
        Blacksburg, VA, USA \\
        0000-0002-8307-8844}
    \and
    \IEEEauthorblockN{2\textsuperscript{nd} Bipasha Banerjee}
    \IEEEauthorblockA{
        \textit{University Libraries} \\
        \textit{Virginia Tech}\\
        Blacksburg, VA, USA \\
        0000-0003-4472-1902}
    \and
    \IEEEauthorblockN{3\textsuperscript{rd} Edward A. Fox}
    \IEEEauthorblockA{
        \textit{Department of Computer Science} \\
        \textit{Virginia Tech}\\
        Blacksburg, VA, USA \\
        0000-0003-1447-6870}
}

\maketitle

\begin{abstract}
As research institutions increasingly commit to supporting the United Nations’ Sustainable Development Goals (SDGs), there is a pressing need to accurately assess their research output against these goals.
Current approaches, primarily reliant on keyword-based Boolean search queries, conflate incidental keyword matches with genuine contributions, reducing retrieval precision and complicating benchmarking efforts.
This study investigates the application of autoregressive Large Language Models (LLMs) as evaluation agents to identify relevant scholarly contributions to SDG targets in scholarly publications.
Using a dataset of academic abstracts retrieved via SDG-specific keyword queries, we demonstrate that small, locally-hosted LLMs can differentiate semantically relevant contributions to SDG targets from documents retrieved due to incidental keyword matches, addressing the limitations of traditional methods.
By leveraging the contextual understanding of LLMs, this approach provides a scalable framework for improving SDG-related research metrics and informing institutional reporting.

\end{abstract}


\section{Introduction}

As universities compete on an international scale, rankings tied to the United Nations’ Sustainable Development Goals (SDGs) offer a common metric to compare social impact.
Rankings affect funding, reputation, and recruitment, so universities often see tracking SDG-related research output as essential to maintaining competitiveness. 
Current methods for identifying an institution's SDG-related research output rely primarily on keyword-based Boolean search queries applied to academic databases, a practice that often allows superficially relevant papers to appear aligned with SDGs without contributing meaningfully to SDG targets.
This imprecision complicates benchmarking efforts, potentially diluting institutional reputations and diminishing the comparability of SDG-related research metrics between institutions.

Recent studies, such as~\cite{armitage_mapping_2020,lemarchand_2022_computational,Heikkil2021}, highlight the limitations of Boolean search strategies in identifying SDG-related research, noting their inability to account for the context or substantive contributions of retrieved documents, which can lead to false positives and missed opportunities to identify impactful research. 
Query expansion using LLMs, as demonstrated by~\cite{bergeron2023ula}, mitigates the narrowness of traditional keyword searches by generating semantically relevant terms. 
However, this approach still operates within the confines of the keyword-based paradigm, leaving unresolved the inherent limitations such as context insensitivity and the inability to discern substantive contributions from superficial mentions.
Multi-label SDG classification is explored in~\cite{yin2024evaluatingSDG} by comparing six LLMs against GPT-4o, with an emphasis on minimizing false positives. 
Their findings highlight the general capability of LLMs to perform reliable SDG tagging. 
Additionally, \cite{garigliotti2024sdg} explores the application of Retrieval-Augmented Generation (RAG) frameworks, built on LLMs such as Llama2 and GPT-3.5, to identify textual passages in environmental reports that align with specific SDG targets. 
These studies demonstrate advances in using LLMs and other machine learning methods for SDG classification, but their reliance on broad relevance criteria and limited exploration of alignment with specific SDG targets leaves open the challenge of distinguishing substantive contributions to these targets, a gap addressed by our study.

\begin{figure}[t]
    \centering
    \begin{tikzpicture}[
        node distance=.5cm and .5cm,
        box/.style={draw, thin, align=center, minimum width=1.5cm, minimum height=1cm, font=\scriptsize},
        arrow/.style={-Stealth, thin}
    ]
    
    \node[box] (database) {Database\\ (Scopus)};
    \node[below=of database, box] (retrieval) {Retrieval\\Agent};
    
    \node[box, right=of database, xshift=0.15cm] (unfiltered) {Unfiltered\\Results};
    
    \node[box, below=of unfiltered] (critic) {Evaluation\\Agent};
    
    \node[box, right=of critic, xshift=0.15cm, yshift=1.0cm] (filtered) {Filtered\\Results};
    
    \draw[arrow] (database) -- (retrieval);
    \draw[arrow] (retrieval) -- (unfiltered);
    \draw[arrow] (unfiltered) -- (critic);
    \draw[arrow] (critic) -- (filtered);
    
    \end{tikzpicture}
    \caption{Absracts retrieved from Scopus via keyword-based queries are classified by an LLM-driven Evaluation Agent, which distinguishes between abstracts with actual contributions to SDG targets and those with only SDG-related terms.}
    \label{fig:architecture}
\end{figure}
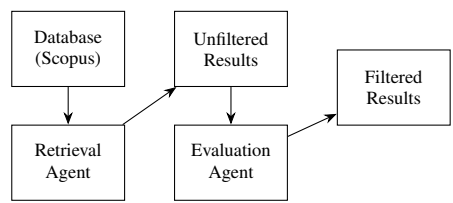

In this study, we propose an alternative approach (see Fig.~\ref{fig:architecture}) that uses an AI \textit{evaluation agent} to distinguish between abstracts with actual contributions to SDG targets from those with only surface-level occurrences of SDG-related terms.
The evaluation agent is built upon an LLM specifically prompted to evaluate abstracts for substantive relevance to SDG targets.
It does so by following structured guidelines that distinguish abstracts that describe concrete contributions, such as measurable actions or findings related to SDG targets, from abstracts that contain SDG-related terms without describing actual contributions to these targets.
This approach represents an application of data science and big data text analytics to scholarly bibliographic data, using small, locally-hosted LLMs to process large datasets with a nuanced understanding of language and context, addressing a critical gap in traditional keyword-based retrieval methods.
The generative capabilities of small, locally hosted LLMs enable us to improve the precision of searches without sacrificing scale, moving beyond simple keyword matching to incorporate contextual understanding and interpretation of SDG targets by capturing subtle nuances that traditional keyword-based methods frequently miss.
See Fig.~\ref{fig:architecture} for a conceptual overview.

\section{Methodology}
To test our approach, we retrieve an initial data set of journal article and conference proceeding abstracts from Scopus, 20,000 for each of the 17 SDGs, using search queries developed and refined by Elsevier's SDG Research Mapping Initiative~\cite{jayabalasingham2019identifying}.
These queries were chosen based on their compatibility with Scopus, established quality and reliability, widespread adoption, and public availability. 
For example, Times Higher Education (THE) uses the Elsevier SDG mapping for its rankings~\cite{the_impact_2024}.

\begin{figure}[t]
    \centering
    \includegraphics[width=0.9\linewidth]{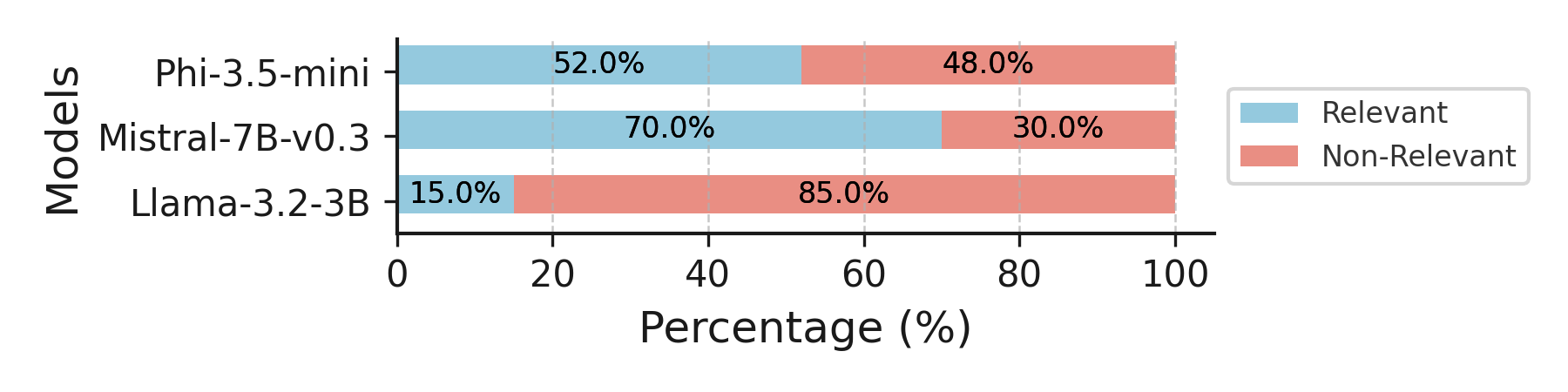}
    \caption{Classification outcomes for Phi-3.5-mini, Mistral-7B-v0.3, and Llama-3.2-3B, showing the percentages of abstracts classified as `Relevant’ or `Non-Relevant’ to SDG 1. Variability across models highlights differences in classification tendencies, suggesting further prompt refinement.}
    \label{fig:init-results}
\end{figure}

The preprocessing steps included removing copyright statements from the abstracts using regular expressions, and excluding documents with missing titles or abstracts to ensure a consistent and usable data set. 
Some documents appeared in the retrieval sets for multiple queries due to the interconnected nature of the SDGs themselves, resulting in queries for different goals sometimes sharing the same keywords. 
Abstracts containing such shared terms were included in the retrieval sets for all relevant queries and these documents were assigned all of the corresponding SDG labels. 
This approach preserved the interdisciplinary nature of the SDGs and allowed the evaluation agent to independently assess each document for relevance to each labeled goal.


In implementations of the evaluation agent, we contrast the use of three LLMs,
chosen for their small memory footprint, local hosting capability, and large context windows: Microsoft's Phi-3.5-mini-instruct~\cite{abdin2024phi3}, Mistral-7B-Instruct-v0.3~\cite{jiang2023mistral}, and Meta's Llama-3.2-3B-Instruct~\cite{dubey2024llama}.
Each model was implemented using an instruction-based prompt that guided the classification of abstracts according to their alignment with specific SDG targets. 

\begin{table}[t]
\centering
\caption{Prompt components used by the evaluation agent}
\label{tab:prompt_breakdown}
\begin{tabular}{@{}p{1.7cm}p{6cm}@{}}
\toprule
\textbf{Component}          & \textbf{Details}\\ 
\midrule
\raggedright \textbf{System Role}        &  Define the task: distinguish between superficial mentions and substantive SDG contributions. \\ 
\addlinespace
\raggedright \textbf{SDG Definition} & SDG 1: End poverty in all its forms everywhere.                                           \\ 
\addlinespace
\raggedright \textbf{Classification Guidelines} &  Relevant: Contributions to SDG targets (e.g., Target 1.2, Target 1.5). Non-Relevant: Superficial mentions. \\ 
\addlinespace
\raggedright \textbf{Example Abstracts}  &  Relevant: Microfinance program increasing economic resilience. Non-Relevant: General economic discussion. \\ 
\addlinespace
\raggedright \textbf{Output Requirements} & Binary classification (Relevant or Non-Relevant) with reasoning referencing abstract text.   \\ 
\bottomrule
\end{tabular}
\end{table}

The prompt consists of a task-specific instruction that defines the evaluation task and emphasizes the distinction between superficial mentions and substantive contributions. 
See Table~\ref{tab:prompt_breakdown} for a summary breakdown of the prompt components. 
The prompt also provides a detailed list of the SDG targets, such as reducing by half the proportion of people living in poverty (SDG 1, Target 1.2) and building the resilience of the poor to climate-related extreme events (SDG 1, Target 1.5). 
These targets serve as criteria for determining relevance. 
Finally, the prompt requires the agent to produce binary classifications of ``Relevant" or ``Non-Relevant" and to provide reasoning that references specific textual evidence from the abstract to justify the classification.

\begin{figure}[t]
    \centering

    \begin{subfigure}[b]{0.9\columnwidth}
        \centering
        \includegraphics[width=\textwidth]{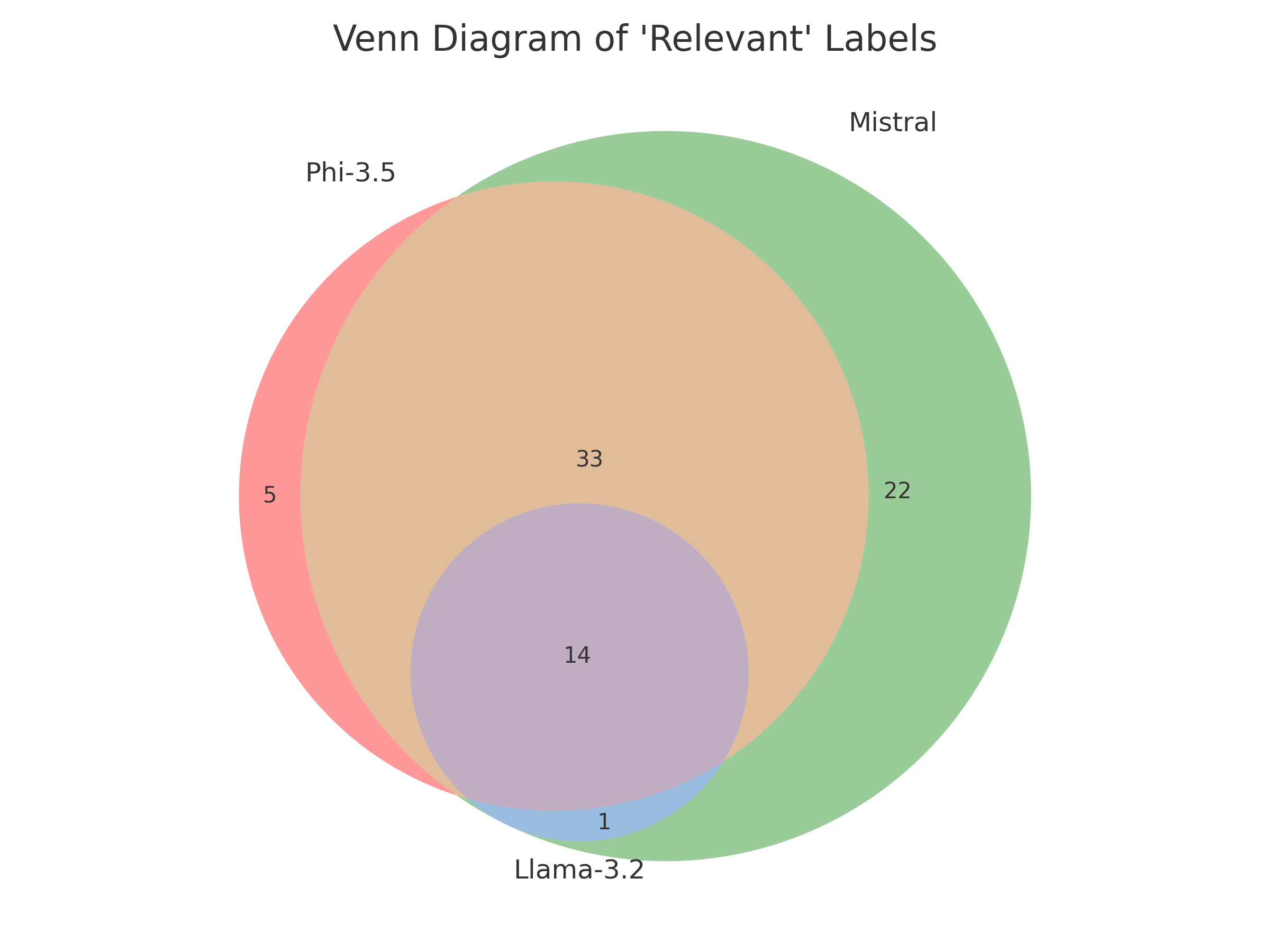} 
        \caption{Agreement among models for \textit{Relevant} classification.}
        \label{fig:relevant_venn}
    \end{subfigure}
    
    \vspace{0.5cm} 
    
    \begin{subfigure}[b]{0.9\columnwidth}
        \centering
        \includegraphics[width=\textwidth]{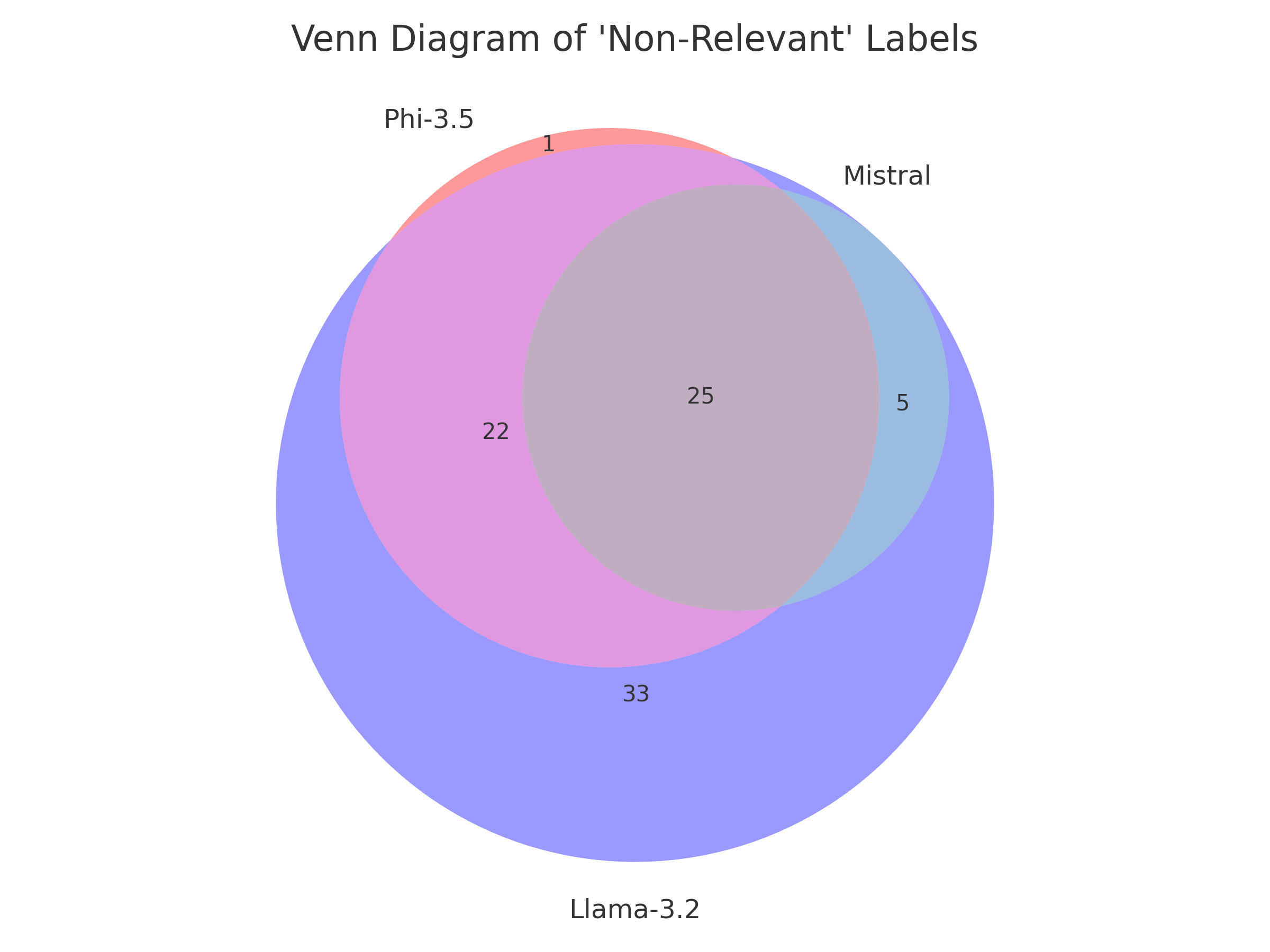} 
        \caption{Agreement among models for \textit{Non-Relevant} classification.}
        \label{fig:non_relevant_venn}
    \end{subfigure}
    
    \caption{Venn diagrams illustrating model agreement for \textit{Relevant} and \textit{Non-Relevant} classifications across the three LLMs (Phi-3.5-mini, Mistral-7B-v0.3, and Llama-3.2-3B).}
    \label{fig:agreement_venn}
\end{figure}

\section{Results}
Our initial results are presented in Fig.~\ref{fig:init-results}.
They indicate clear differences in how each LLM interprets relevance. 
We analyzed proportions of abstracts labeled `Relevant' versus `Non-Relevant' and inter-model agreement rates.
Phi-3.5-mini labeled 52\% of abstracts as `Relevant' and 48\% as `Non-Relevant'. showing relatively balanced outputs. 
Mistral-7B was more expansive, assigning 70\% of abstracts to the `Relevant' category, while Llama-3.2 was more selective, labeling only 15\% as `Relevant'. 
These variations suggest differences in the way each model applies the evaluation criteria outlined in the prompt.

Pairwise agreement rates reveal distinct patterns, as illustrated in the Venn diagrams in Fig.~\ref{fig:agreement_venn}. 
In the `Relevant' classification, Phi-3.5-mini and Mistral-7B exhibit the largest overlap, consistent with their shared tendency to label more abstracts as `Relevant'. 
However, the overlap between Llama-3.2 and either of the other models is minimal, reflecting its stricter filtering criteria. 
The small intersection of all three models for `Relevant' classifications underscores the variability in their application of the prompt.
Conversely, the `Non-Relevant' Venn diagram demonstrates higher alignment across the models, with a substantial proportion of abstracts classified as `Non-Relevant' by all three. 

The observed variations may reflect the design and emphasis of each model. 
Mistral-7B's broader classifications suggest a bias toward including abstracts with indirect or thematic relevance, while Llama-3.2's stricter thresholds likely prioritize explicit contributions. 
Phi-3.5-mini occupies a middle ground, suggesting potential as a mediator in a hybrid classification system.
These findings underscore the value of integrating diverse classification strategies in a multi-agent framework. 
For instance, Mistral-7B could act as a preliminary filter to capture a wide array of abstracts, with Phi-3.5-mini refining classifications and Llama-3.2 providing a final, stricter assessment. 
Such a system could leverage the inclusivity of Mistral-7B while retaining the precision of Llama-3.2, balancing recall and specificity.

Given the differing classifications and agreement rates, an ensemble approach, implemented as a multi-agent conversation, could be a novel solution for strengthening the reliability of relevance classifications.

\section{Conclusion}
This study investigates the utility of small, locally hosted LLMs as evaluation agents for improving the precision of classifying research contributions to SDG targets. 
By addressing the contextual and semantic limitations of traditional keyword-based methods, these models demonstrate an ability to differentiate genuine contributions from superficial mentions within large bibliographic data sets. 
However, some limitations of the study warrant consideration. 
First, our reliance on prompt design introduces sensitivity to variations in phrasing or task framing, which may limit generalizability across datasets or SDGs. 
Second, the models were evaluated using abstracts rather than full-text articles, potentially omitting context necessary for accurate relevance determination. 
Finally, this study focused primarily on SDG 1, leaving the broader applicability across all 17 SDGs for future exploration. 
Our findings reveal divergences in classification thresholds across the tested models, suggesting opportunities to leverage their complementary strengths in ensemble or multi-agent frameworks. 
Future work will explore such frameworks, enabling iterative evaluations and inter-agent feedback to improve consistency and reliability in SDG relevance classification with LLMs. 

\bibliography{ref}
\bibliographystyle{IEEEtran}
\end{document}